\documentclass[10pt,twocolumn,letterpaper]{article}

\usepackage{cvpr}
\usepackage{times}
\usepackage{epsfig}
\usepackage{graphicx}
\usepackage{amsmath}
\usepackage{amssymb}
\usepackage{subcaption}
\usepackage{acro}
\DeclareAcronym{MRI}{
short=MRI,
long=Magnetic Resonance Imaging
}
\DeclareAcronym{CT}{
short=CT,
long= Computed Tomography
}
\DeclareAcronym{CS}{
short=CS,
long=Compressed Sensing
}
\DeclareAcronym{PI}{
short=PI,
long=Parallel Imaging
}
\DeclareAcronym{CNN}{
short=CNN,
long=Convolutional Neural Network
}
\DeclareAcronym{TR}{
short=TR,
long=Repetition Time
}
\DeclareAcronym{TE}{
short=TE,
long=Echo Time
}
\DeclareAcronym{T1WI}{
short=T1WI,
long=T1 Weighted Imaging
}
\DeclareAcronym{T2WI}{
short=T2WI,
long=T2 Weighted Imaging
}
\DeclareAcronym{FLAIR}{
short=FLAIR,
long=Fluid-Attenuated Inversion Recovery
}
\DeclareAcronym{FT}{
short=FT,
long=Fourier Transform
}
\DeclareAcronym{IFT}{
short=IFT,
long=Inverse Fourier Transform
}

\def\@fnsymbol#1{\ensuremath{\ifcase#1\or *\or \dagger\or \ddagger\or
   \mathsection\or \mathparagraph\or \|\or **\or \dagger\dagger
   \or \ddagger\ddagger \else\@ctrerr\fi}}
\newcommand{\ssymbol}[1]{^{\@fnsymbol{#1}}}

\newcommand\Tstrut{\rule{0pt}{2.0ex}}         
\newcommand\Bstrut{\rule[-0.9ex]{0pt}{0pt}}   

\cvprfinalcopy 

\ifcvprfinal\fi
\begin{document}

\title{DuDoRNet: Learning a Dual-Domain Recurrent Network for Fast MRI Reconstruction with Deep T1 Prior}

\author{Bo Zhou$^1$\\
$^1$Department of Biomedical Engineering,\\
Yale University\\
{\tt\small bo.zhou@yale.edu}
\and
S. Kevin Zhou$^{2,3}$\\
$^2$Chinese Academy of Sciences\\
$^3$Peng Cheng Laboratory, Shenzhen\\
{\tt\small s.kevin.zhou@gmail.com}
}

\maketitle

\begin{abstract}
\ac{MRI} with multiple protocols is commonly used for diagnosis, but it suffers from a long acquisition time, which yields the image quality vulnerable to say motion artifacts. To accelerate, various methods have been proposed to reconstruct full images from under-sampled k-space data. However, these algorithms are inadequate for two main reasons. Firstly, aliasing artifacts generated in the image domain are structural and non-local, so that sole image domain restoration is insufficient. Secondly, though MRI comprises multiple protocols during one exam, almost all previous studies only employ the reconstruction of an individual protocol using a highly distorted undersampled image as input, leaving the use of fully-sampled short protocol (say T1) as complementary information highly underexplored. In this work, we address the above two limitations by proposing a Dual Domain Recurrent Network (DuDoRNet) with deep T1 prior embedded to simultaneously recover k-space and images for accelerating the acquisition of MRI with a long imaging protocol. Specifically, a Dilated Residual Dense Network (DRDNet) is customized for dual domain restorations from undersampled MRI data. Extensive experiments on different sampling patterns and acceleration rates demonstrate that our method consistently outperforms state-of-the-art methods, and can reconstruct high quality MRI.
\end{abstract}

\section{Introduction}
\begin{figure*}[htb!]
\centering
\includegraphics[width=1.00\textwidth]{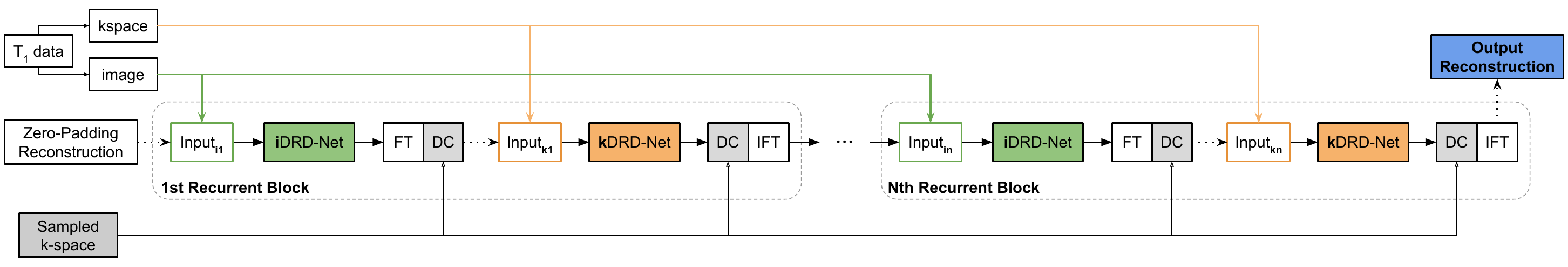}
\caption{The architecture of our proposed \textbf{Du}al \textbf{Do}main \textbf{R}ecurrent \textbf{Net}work (\textbf{DuDoRNet}). DC denotes the data consistency layer. Each recurrent block contains one network for image domain restoration and one network for k-space domain restoration with two interleaved DC.}
\label{fig:pipline}
\end{figure*}

\ac{MRI} is one of the most applied imaging procedures for disease diagnosis and treatment planning. As a non-invasive, radiation-free, and in-vivo imaging modality, it provides significantly better soft-tissue contrast than many other imaging modalities and offers accurate measurements of both anatomical and functional signals. However, its long acquisition time, owing to sampling full k-space, especially under the multiple protocols requiring long \ac{TE} and \ac{TR}, could lead to significant artifacts in the reconstructed image caused by patient or physiological motions during acquisitions, \ie cardiac motion and respiration. Furthermore, a long acquisition time also limits the availability of MR scanner for patients and causes delayed patient care in the medical system. 

To accelerate the \ac{MRI} acquisition, various efforts have been made for improving the reconstruction image quality with undersampled k-space data. Previously, \ac{CS} and Parallel Imaging have achieved significant progresses in fast \ac{MRI} \cite{otazo2010combination,griswold2002generalized,huang2005k}. In \ac{CS}-\ac{MRI}, one assumes that images have a sparse representation in certain domains \cite{lustig2007sparse,murphy2012fast,liang2009accelerating}. In conventional \ac{CS}-\ac{MRI}, previous works focus on using sparse coefficients in sparsifying transforms, \ie wavelet transform \cite{qu2010combined,zhang2015exponential} and contourlet transform \cite{qu2010iterative}, combined with regularization parameters, \ie total variation, to solve the ill-posed inverse problems in an iterative manner. However, these iterative minimization process based on sparsifying transforms tends to generate smaller sparse coefficient values, and lead to loss of details and unwanted artifacts in the reconstruction when undersampling rate is high \cite{ravishankar2010mr}. Thus, current \ac{CS}-\ac{MRI} is limited to a rate of $2 \sim 3$ \cite{lustig2007sparse,ravishankar2010mr}. Furthermore, reconstruction based on these iterative algorithms is time-consuming; thus it is challenging to deploy them in near real-time MRI scenarios, \ie Cardiac-\ac{MRI} and Functional-\ac{MRI}. More recently, triggered by the success of computer vision \cite{lecun2015deep}, deep learning based algorithms have been developed for fast \ac{MRI} reconstruction and demonstrated significant advantages \cite{wang2016accelerating,schlemper2017deep,sun2016deep,hammernik2018learning,liu2019theoretically,lonning2019recurrent,han2019k,mardani2018deep,zhang2019reducing,zhu2018image}. Wang \etal \cite{wang2016accelerating} first proposed to train a multi-layer \ac{CNN} to recover the fully sampled MRI image from undersampled MRI image using supervised training with paired data. Similarly, Jin \etal \cite{jin2017deep} proposed to use U-Net \cite{ronneberger2015u} to solve the inverse problem in imaging. Yang \etal \cite{yang2017dagan} developed a De-Aliasing GAN that uses a U-Net as the generator with a loss function consisting of four components: an image domain loss, a frequency domain loss, a perceptual loss, and an adversarial loss. Quan \etal \cite{quan2018compressed} introduced a RefineGAN that uses a U-Net structure based generator with a cyclic loss for MRI de-aliasing. However, individual training is required for different sampling patterns and undersampling rates. Schlemper \etal \cite{schlemper2017deep,qin2018convolutional} came up with the data consistency layer in a deep cascade of \ac{CNN}, which ensures the consistency of the reconstruction image in k-space and potentially reduces the issue of overfitting in deep training. Other than learning from pre-defined undersampling patterns, Zhang \etal \cite{zhang2019reducing} proposed to use active learning to determine the 1D Cartesian undersampling pattern on the fly during the \ac{MRI} acquisition. These pioneering deep learning based methods surpass conventional \ac{CS} algorithms owing to the high-nonlinearity properties of data-driven feature extraction and achieve significantly lower computation during run-time. However, existing deep learning based methods have three major limitations. In this work, we aim to break these limitations.

Firstly, the above-mentioned algorithms are principally learned in the image domain alone, with a few amendments that use the frequency domain in the loss design \cite{yang2017dagan} or in the data consistency layer \cite{schlemper2017deep}. All of these deep learning based algorithms are designed to receive an image reconstructed from undersampled k-space as input and output an image as if reconstructed from fully sampled k-space; but unfortunately in the input images to the CNNs, detailed structures are likely distorted or even disappear. While there are recent attempts on restoring fully sampled image from undersampled k-space \cite{eo2018kiki}, the large scale trainable parameters limits the model can only be trained in an incremental fashion which could be further improved. As the first step to tackle this issue, we develop a dual domain learning scheme in MRI, which allows the network to restore the data in both image and frequency domains in a recurrent fashion. Previous work on CT metal artifact reduction also demonstrated the advantages of cross domain learning \cite{lin2019dudonet}.

Secondly, the previous studies are limited to conventional network design, such as multi-layer \ac{CNN} and U-Net, and there are few attempts to design a customized network structure for undersampled \ac{MRI} reconstruction. Inspired by the recent development of super-resolution imaging techniques \cite{dong2015image,kim2016accurate,zhang2018residual,hu2018squeeze}, we propose a Dilated Residual Dense Network (DRD-Net) with a Squeeze-and-Excitation Dilated Dense Residual Block (SDRDB) as the building module. The DRD-Net is used for both image and frequency domain restorations. Our SDRDB is customized for \ac{MRI} reconstruction task. In fast MRI acquisitions, we observe significant sparsity in undersampled k-space, especially when undersampling rate is high. Previous studies have demonstrated that better k-space recovery can be achieved via using non-local information interpolation, such as GRAPPA \cite{griswold2002generalized}. Based on this assumption, the first motivation of our DRD Block design in k-space is synthesis of missing k-space from a large receptive field by utilizing non-local k-space data, thus bringing more robustness and reliability. In the image domain, human organ anatomy is correlated in different regions. Our DRD Block with a large receptive field in the image domain can better capture this correlation between anatomical regions and synthesize the missing anatomical information even when the signal is highly distorted. 

Thirdly, previous studies have not fully explored the use of \ac{MRI} protocol that requires a short acquisition time as deep prior to guide the \ac{MRI} reconstruction process. In clinical routines, typical total scanning time for \ac{T1WI}, \ac{T2WI}, and \ac{FLAIR} is $\sim 20$ mins, in which \ac{T2WI} and \ac{FLAIR} take the majority due to their long \ac{TR} and \ac{TE}. However, using the undersampled image with the detailed structures may already disappeared as input, the existing methods could synthesize artificial structure that does not belong to the patient. Recently, Xiang \etal explored the merits of using \ac{T1WI} as additional channel input in the image domain to aid \ac{T2WI} reconstruction \cite{xiang2018ultra}. To further address this issue, we propose to use \ac{T1WI} as deep prior in both image domain and k-space domain for improving the \ac{MRI} reconstruction fidelity, given that the structural information in \ac{T1WI} is highly correlated with that in different \ac{MRI} protocols.

In summary, we propose a \textbf{Du}al \textbf{Do}main \textbf{R}ecurrent \textbf{Net}work (DuDoRNet) embedding with T1 priors to address these problems by learning two DRD-Nets on dual domains in a recurrent fashion to restore k-space and image domains simultaneously. Our method (Figure \ref{fig:network}) consists of three major parts: recurrent blocks comprised of image restoration network (iDRD-Net) and k-space restoration network (kDRD-Net), recurrent T1 priors embedded in image and k-space domains, and recurrent data consistency regularization. A correct reconstruction should ensure the consistency in both domains linked by the linear operation of \ac{FT}. Our intuition is that image domain restoration can be enhanced by fusing signal back-propagated from the k-space restoration, vice versa. Given sparse signal in both domains, DRD-Net with a large receptive field can sense more signal for a better restoration. Our recurrent learning can better avoid overfitting in directly optimizing restoration networks in dual domains. Extensive experiments on MRI patients with different sampling patterns and acceleration rates demonstrate that our DuDoRNet generates superior reconstructions.

\section{Problem Formulation}
Denoting a 2D k-space with complex values as $k$, and a 2D image reconstructed from $k$ as $x$, we need to reconstruct a fully sampled image $x$ from both the undersampled k-space ($k_u$) and reconstructed image ($x_u$). The relationship between $x$ and $k$ can be written as:
\begin{equation}\small
    k_u = M \odot k_f = M \odot \mathcal{F}(x_f) ,
\end{equation}
\begin{equation}\small
    x_u = \mathcal{F}^{-1} (k_u) = \mathcal{F}^{-1} (M \odot \mathcal{F}(x_f)) ,
\end{equation}
where $M$ represents the binary undersampling mask used for accelerations; $k_f$ and $k_u$ denote the fully sampled and the undersampled k-space, respectively; $x_f$ and $x_u$ denote the images reconstructed from fully sampled and undersampled k-space, respectively; $\odot$ is the element-wise multiplication operation; and $\mathcal{F}$ and $\mathcal{F}^{-1}$ are \ac{FT} and \ac{IFT}. 

To tackle the ill-posed inverse problem of reconstructing $x_f$ from limited sampled data, we propose to restore both image domain and k-space domain. In image domain restoration, the optimization can be expressed as minimizing the prediction error:
\begin{equation} \small \label{eqn:img_opt_x}
    \underset{\tilde x}{\arg\min} || x_f - \tilde x ||^2_2 = \underset{\theta_x}{\arg\min} || x_f - \mathcal{P}_x (x_u;\theta_x) ||^2_2 ,
\end{equation}
where $\tilde x$ is the prediction of the fully sampled image ($x_f$) generated by the estimation function ($\mathcal{P}_x$) with its parameters ($\theta_x$) using the undersampled image ($x_u$) as input. Furthermore, the data consistency constraint \cite{schlemper2017deep} is often used in addition to the prediction error:
\begin{subequations} \small
\begin{align}
    \underset{\theta_x}{\arg\min} &( || x_f - \mathcal{P}_x (x_u;\theta_x) ||^2_2 \label{eqn:img_opt_dc1}\\
                                  & + \lambda ||k_u - M \odot \mathcal{F} (\mathcal{P}_x (x_u;\theta_x)) ||^2_2 ), \label{eqn:img_opt_dc2}
\end{align}
\end{subequations}
where \eqref{eqn:img_opt_dc1} is the same as \eqref{eqn:img_opt_x}, and \eqref{eqn:img_opt_dc2} is the regularization term for data consistency. Similarly, we can formulate the optimization target for k-space restoration as:
\begin{equation} \small \label{eqn:img_opt_k}
    \underset{\tilde k}{\arg\min} || k_f - \tilde k ||^2_2 = \underset{\theta_k}{\arg\min} || k_f - \mathcal{P}_k (k_u;\theta_k) ||^2_2 ,
\end{equation}
where $\tilde k$ is the prediction of the fully sampled k-space ($k_f$) generated by the estimation function ($\mathcal{P}_k$) with its parameters ($\theta_k$) using the undersampled k-space ($k_u$) as input. Combining the image domain and k-space domain restoration with data consistency, the target function can thus be formulated as:
\begin{subequations} \small
\begin{align}
    \underset{\theta_k,\theta_x}{\arg\min} ( & || k_f - \mathcal{P}_k (\mathcal{F}(\mathcal{P}_x (x_u;\theta_x));\theta_k) ||^2_2 \\
     + & || x_f - \mathcal{P}_x (\mathcal{F}^{-1}(\mathcal{P}_k (k_u;\theta_k));\theta_x) ||^2_2 \\
     + \lambda & || k_u - M \odot \mathcal{F} (\mathcal{P}_x (\mathcal{F}^{-1} (\mathcal{P}_k (k_u;\theta_k)) ; \theta_x)) ||^2_2),
\end{align}
\end{subequations}
Directly optimizing multiple terms in the above target function is challenging in traditional network design, owing to its high computational complexity, overfitting, and local optima problems. Thus, we propose a dual-domain recurrent learning strategy that optimizes $\theta_x$ and $\theta_k$ recurrently. Our proposed approach is illustrated in details in the following sections.

\section{Methods}
The overall pipeline of our network is illustrated in Figure \ref{fig:pipline}. It consists of three major parts: 1) the dual domain recurrent network consisting of recurrent blocks with image and k-space restoration networks in it; 2) the deep prior information generated from T1 data in the image and k-space domains for feeding into the recurrent blocks of the network; and 3) the recurrent data consistency regulations using sampled k-space data. In each recurrent block, we propose to use a Dilated Residual Dense Network (DRD-Net) for both image and k-space restorations.

\subsection{Dilated Residual Dense Network}
\begin{figure}[!htb]
\centering
\includegraphics[width=0.46\textwidth]{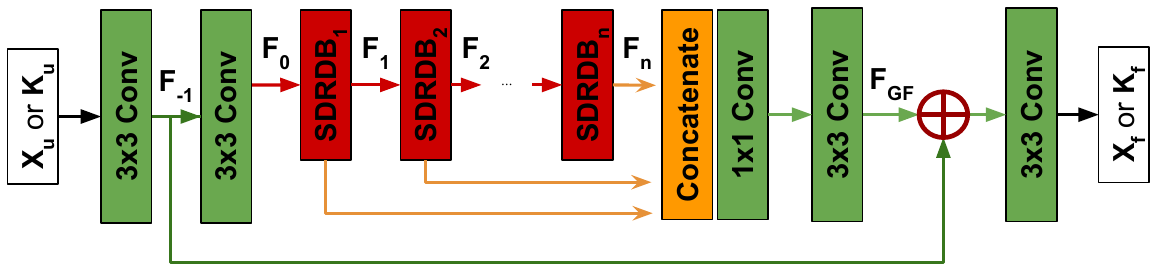}
\caption{The architecture of our \textbf{D}ilated \textbf{R}esidual \textbf{D}ense Network (\textbf{DRD-Net}) with building modules of SDRDB shown in Figure \ref{fig:network_block}. The input can be either $x_u$ in image domain or $k_u$ in k-space domain. Convolution operation is followed by ReLU.}
\label{fig:network}
\end{figure}

We develop a network structure for both \ac{MRI} image and k-space restoration, called Dilated Residual Dense Network (DRD-Net). The idea is to use a Squeeze-and-excitation Dilated Residual Dense Block (SDRDB) as the backbone in our DRD-Net. The DRD-Net design and SDRDB are shown in Figure \ref{fig:network} and Figure \ref{fig:network_block}, respectively. Compared to RDN \cite{zhang2018residual}, we customize the local and global structure design for our \ac{MRI} reconstruction task. 

\subsubsection{Global Structure}
Our DRD-Net consists of three parts: initial feature extraction (IFE) via two sequential $3 \times 3$ convolution layers, multiple SDRDBs followed by global feature fusion, and global residual learning. The overall pipeline is as follow:
\begin{equation} \small
    F_{-1} = \mathcal{P}_{IFE_{1}} (X_u) ,
\end{equation}
\begin{equation} \small
    F_{0} = \mathcal{P}_{IFE_{2}} (F_{-1}) ,
\end{equation}
where $\mathcal{P}_{IFE_{1}}$ and $\mathcal{P}_{IFE_{2}}$ denote the first and second convolutional operations in IFE, respectively. The first extracted feature $F_{-1}$ is used for global residual learning in the third part. The second extracted feature $F_{0}$ is used as SDRDB input. If there are $n$ SDRDBs, the $n$-th output $F_{n}$ can be written as:
\begin{equation} \small \label{eq:sdrdb}
    F_{n} = \mathcal{P}_{SDRDB_{n}} (F_{n-1}) , 
\end{equation}
where $\mathcal{P}_{SDRDB_{n}}$ represents the n-th SDRDB operation with $n \geq 1$. Given the extracted local features from a set of SDRDBs, we apply global feature fusion (GFF) to extract the global feature via: 
\begin{equation} \small
    F_{GF} = \mathcal{P}_{GFF} (\{F_{1}, F_{2}, \dots, F_{n}\}) ,
\end{equation}
where $\{ \}$ denotes the concatenation operation along feature channel. Our global feature fusion function $\mathcal{P}_{GFF}$ consists of a $1 \times 1$ and $3 \times 3$ convolution layers to fuse the extracted local features from different levels of SDRDB. The GFF output is used as input for global residual learning:
\begin{equation} \small
    X_f = \mathcal{P}_{final} (F_{GF} + F_{-1}) ,
\end{equation}
The element-wise addition of global feature and initial feature are fed into our final $3 \times 3$ convolution layer for reconstruction output.

\subsubsection{SDRD Block}
\begin{figure}[htb!]
\centering
\includegraphics[width=0.48\textwidth]{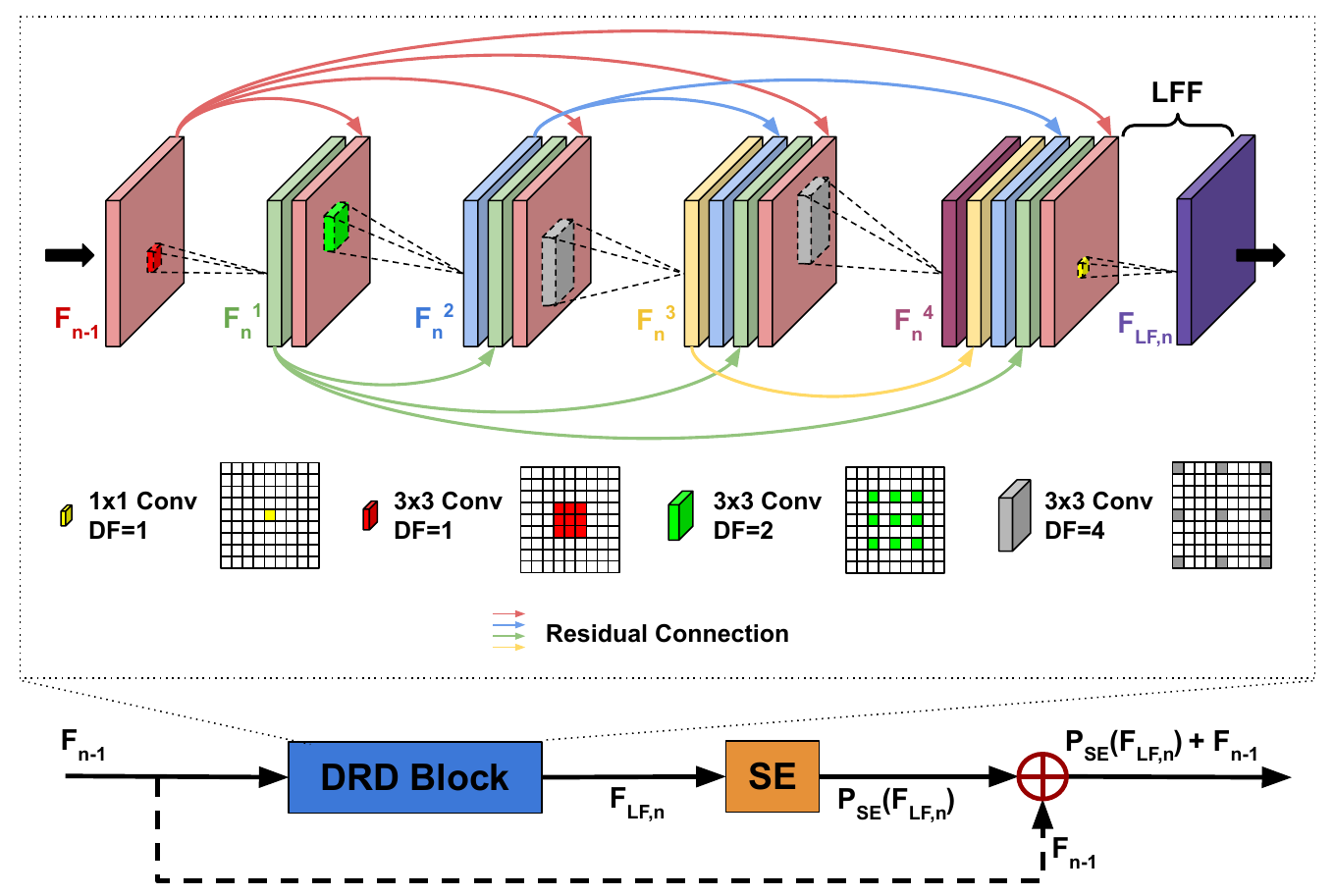}
\caption{The structure of our \textbf{S}queeze-and-excitation \textbf{D}ilated \textbf{R}esidual \textbf{D}ense Block (\textbf{SDRDB}).}
\label{fig:network_block}
\end{figure}

\begin{figure}[!htb]
\centering
\includegraphics[width=0.41\textwidth]{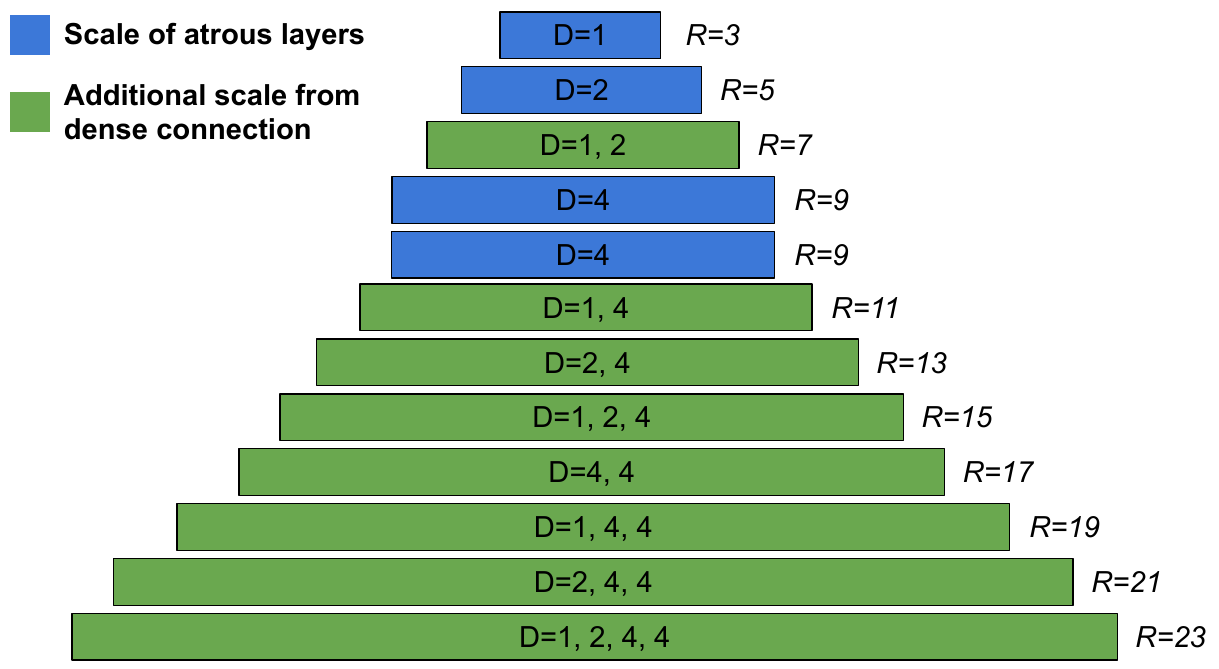}
\caption{Illustration of the scale pyramid using our densely connected atrous layers with kernel size of $3 \times 3$ and dilation factors of 1,2,4, and 4. Our densely connection setting creates diverse kernel sizes with much larger receptive fields (blue + green) than naive sequential dilated layers (blue only). R denotes the size of receptive field created from the dilation (D) combination in each block.}
\label{fig:pyramid}
\end{figure}

The design of our SDRDB is shown in Figure \ref{fig:network_block}. It contains four densely connected atrous convolution layers \cite{chen2017deeplab}, local feature fusion, Squeeze-and-Excitation (SE) \cite{hu2018squeeze}, and local residual learning. Expanding the expression of SDRDB, the $t$-th convolution output in $n$-th SDRDB can be written as:
\begin{equation} \small
    F_{n}^{t} = \mathcal{H}_{n}^{t} \{F_{n-1}, F_{n}^{1}, \dots ,F_{n}^{t-1}\} ,
\end{equation}
where $\mathcal{H}_{n}^{t}$ denotes the $t$-th convolution followed by Leaky-ReLU in the $n$-th SDRDB, $\{ \}$ is the concatenation operation along feature channel, and the number of convolution is set to $t \leq 4$. Our SDRDB begins by composing a feature pyramid using 4 atrous convolution layers with dilation rates of 1, 2, 4, and 4. For an atrous convolution layer with kernel size (K) and kernel dilation (D), the receptive field (R) can be written as $R = K + (K-1) \times (D-1)$. The combination of two atrous convolution layers creates a new receptive field of $R_{comb} = R_1 + R_2 -1$. With that being said, the dense connection over 4 atrous layers enables diverse combinations of layers with various receptive fields, which can more efficiently extract features from different scales than traditional dilation approaches. Figure \ref{fig:pyramid} demonstrates the feature pyramid from our 4 densely connected atrous convolution layers. 

\begin{table*} [htb!]
\footnotesize
\centering
\caption{Quantitative comparison of T2 reconstructions from different undersampling patterns and methods at R = 5. Best results with and without T1 prior are marked in \textcolor{red}{red} and \textcolor{blue}{blue}, respectively.}
\label{tab:scan_metrics}
    \begin{tabular}{|c||c|c|c|c|c|c|c||c|c|}
        \hline
        \textbf{Cartesian} & ZP & GRAPPA\cite{griswold2002generalized} & TV\cite{ma2008efficient} & Wang\cite{wang2016accelerating} & DeepCas\cite{schlemper2017deep} & RefGAN\cite{quan2018compressed} & Ours w/o Prior & UF-T2 \cite{xiang2018ultra} & Ours \Tstrut\Bstrut\\
        \hline
        PSNR [dB]          & 22.498 & 23.318 & 23.026 & 26.061 & 26.993 & 25.848 & \textcolor{blue}{27.834} & 30.594 & \textcolor{red}{32.511} \Tstrut\Bstrut\\
        SSIM               & 0.667  & 0.730  & 0.725  & 0.828  & 0.859  & 0.819  & \textcolor{blue}{0.898}  & 0.929  & \textcolor{red}{0.957} \Tstrut\Bstrut\\
        MSE($\times 10^2$) & 0.622  & 0.508  & 0.555  & 0.306  & 0.221  & 0.340  & \textcolor{blue}{0.178}  & 0.097  & \textcolor{red}{0.066} \Tstrut\Bstrut\\
        \hline \hline
        \textbf{Radial} & ZP & GRAPPA & TV & Wang & DeepCas & RefGAN & Ours w/o Prior & UF-T2 & Ours \Tstrut\Bstrut\\
        \hline
        PSNR [dB]          & 24.294 & 29.548 & 27.222 & 32.586 & 34.955 & 35.110 & \textcolor{blue}{37.270} & 33.318 & \textcolor{red}{40.815} \Tstrut\Bstrut\\
        SSIM               & 0.581  & 0.822  & 0.727  & 0.889  & 0.957  & 0.959  & \textcolor{blue}{0.974}  & 0.939  & \textcolor{red}{0.981} \Tstrut\Bstrut\\
        MSE($\times 10^2$) & 0.412  & 0.122  & 0.216  & 0.102  & 0.025  & 0.023  & \textcolor{blue}{0.021}  & 0.049  & \textcolor{red}{0.010} \Tstrut\Bstrut\\
        \hline \hline
        \textbf{Spiral}    & ZP & GRAPPA & TV & Wang & DeepCas & RefGAN & Ours w/o Prior & UF-T2 & Ours \Tstrut\Bstrut\\
        \hline
        PSNR [dB]          & 26.181 & 31.893 & 33.890 & 35.987 & 43.867 & 36.049 & \textcolor{blue}{48.418}  & 37.793 & \textcolor{red}{49.186} \Tstrut\Bstrut\\
        SSIM               & 0.675  & 0.882  & 0.909  & 0.932  & 0.972  & 0.960  & \textcolor{blue}{0.991}   & 0.961  & \textcolor{red}{0.993} \Tstrut\Bstrut\\
        MSE($\times 10^2$) & 0.266  & 0.076  & 0.046  & 0.029  & 0.0046 & 0.024 & \textcolor{blue}{0.0018}  & 0.019  & \textcolor{red}{0.0014} \Tstrut\Bstrut\\
        \hline
    \end{tabular}
\end{table*}

Then, we apply our local feature fusion (LFF), consisting of $1 \times 1$ convolution layer and SE layer, to fuse the output from the last SDRDB and all convolution layers in current SDRDB. Thus, the LFF output can be expressed as:
\begin{equation} \small
    F_{LF,n} = \mathcal{P}_{LFF,n} ( \{F_{n-1}, F_{n}^{1}, F_{n}^{2}, F_{n}^{3} ,F_{n}^{4}\} ) ,
\end{equation}
where $\mathcal{P}_{LFF,n}$ denotes the LFF operation. Finally, we apply SE and the local residual learning to LFF output by adding the residual connection from SDRDB input. Thus, the SDRDB output is:
\begin{equation} \small
    F_{n} = \mathcal{P}_{SE}(F_{LF,n}) + F_{n-1} ,
\end{equation}

\subsection{Dual-Domain Recurrent Learning}
In this section, we present details about our proposed DuDoRNet. As illustrated in Figure \ref{fig:pipline}, each of the recurrent blocks of DuDoRNet contains one image restoration DRD-Net (iDRD-Net), one k-space restoration DRD-Net (kDRD-Net), and two data consistency layers interleaved. The T1 deep prior provides data in both image space and k-space that are fed into DuDoRNet recurrent blocks. In the $n$-th recurrent block, denoting the image input as $x_{u_{n}}$, the image restoration optimization target can be written as:
\begin{subequations} \small
\begin{align}
    \underset{\theta_{iDRD}}{\arg\min} ( & || x_{f} - \mathcal{P}_{iDRD} (x_{u_{n}},x_{T1};\theta_{iDRD}) ||^2_2 \\
    + \lambda & || k_{u_{n}} - M \odot \mathcal{F} (\mathcal{P}_{iDRD} (x_{u_{n}},x_{T1};\theta_{iDRD})) ||^2_2 ) , \nonumber
\end{align}
\end{subequations}
where $\mathcal{P}_{iDRD}$ is the image restoration network based on DRD-Net with parameters $\theta_{iDRD}$. The image output from the last recurrent block $x_{u_{n}}$ and T1 image prior $x_{T1}$ are concatenated channel-wise for inputting into iDRD-Net. $M$ is the binary undersampling mask used for accelerations. The k-space values in $M$ can be altered after inference through iDRD-Net since it only optimizes the first term. To maintain the k-space fidelity at sampled locations $z$ of $M$, we add the data consistency as the second term. Denoting the output of iDRD-Net as $x_{iDRD_{n}}$ and its \ac{FT} as $k_{iDRD_{n}} = \mathcal{F}(x_{iDRD_{n}})$, the corresponding output from data consistency layer can be thus formulated as:
\begin{equation} \small
  k_{iDRD_{n}}=
  \begin{cases}
    \frac{\lambda k_{iDRD_{n}}(z) + k_{u_{n}}(z)}{\lambda + 1} & \text{if $M(z) = 1$} \\[6pt]
    \quad\quad k_{iDRD_{n}}(z) & \text{if $M(z) = 0$}
  \end{cases}
\end{equation}
where $\lambda$ controls the level of linear combination between sampled k-space values and predicted values. When $\lambda=0$, the sampled k-space directly substitutes the prediction at $z$ in k-space. Denoting this output as $k_{u_{n}}=k_{iDRD_{n}}$, then $k_{u_{n}}$ and T1 k-space prior $k_{T1}$ are concatenated channel-wise to feed into the kDRD-Net for k-space restoration. Similarly, the k-space restoration optimization target can be written as:
\begin{subequations} \small \label{eq:img_opt_kdrd}
\begin{align}
    \underset{\theta_{kDRD}}{\arg\min} ( & || k_{f} - \mathcal{P}_{kDRD} (k_{u_{n}},k_{T1};\theta_{kDRD}) ||^2_2 \\
    + \lambda & || k_{u_{n}} - M \odot \mathcal{P}_{kDRD} (k_{u_{n}},k_{T1};\theta_{kDRD})) ||^2_2) , \nonumber
\end{align}
\end{subequations}
where $\mathcal{P}_{kDRD}$ is the k-space restoration network based on DRD-Net with network parameters $\theta_{kDRD}$. Similarly here, the second term is to ensure the data consistency in the restored k-space. Thus, the loss function for each recurrent block is $\mathcal{L}_{i_n} + \mathcal{L}_{k_n}$. The final loss is the summation of recurrent blocks losses: $\sum_{n=1}^{N_{rec}} (\mathcal{L}_{i_n} + \mathcal{L}_{k_n})$, where $N_{rec}$ denotes the number of recurrent blocks.

In our experiments, $\lambda$ is set to $0.01$, $N_{rec}$ is set to $5$, and the number of SDRDB is set to $3$. Each recurrent block in our DuDoRNet share the same network parameters. The final reconstruction output during testing is obtained by applying \ac{IFT} to the last kDRD-Net output after data consistency operation.

\section{Experiments}
\begin{figure*}[htb!]
\centering
\includegraphics[width=0.92\textwidth]{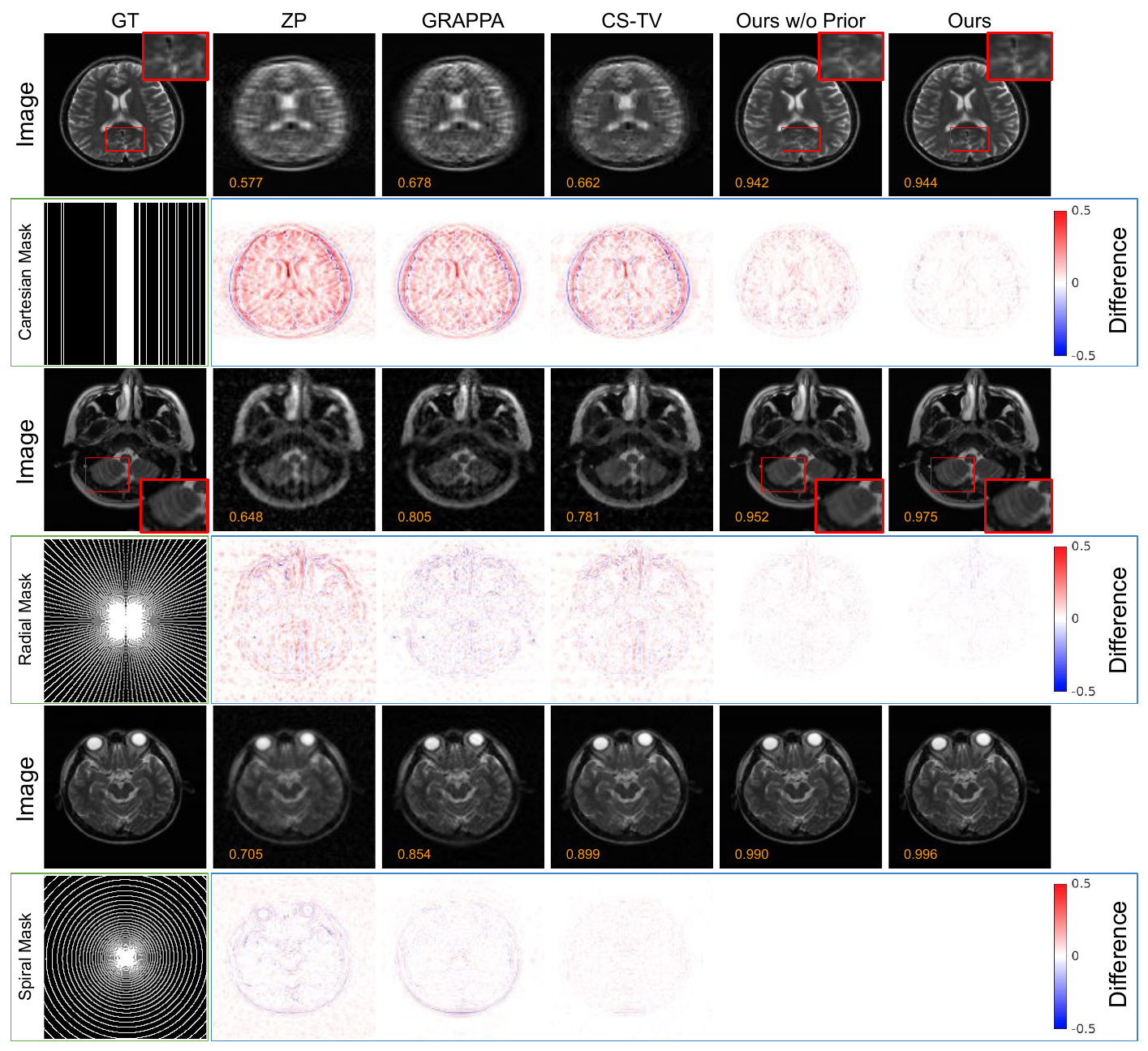}
\caption{Reconstruction results from Cartesian/Radial/Spiral trajectory at under-sampling rate R = 5. The sampling pattern mask and difference images are shown on the second, fourth, and sixth row. Red boxes illustrate the enlarged views on details. The SSIM is indicated on the bottom left of the image.}
\label{fig:comp_image1}
\end{figure*}

\begin{figure}[htb!]
\centering
\includegraphics[width=0.42\textwidth]{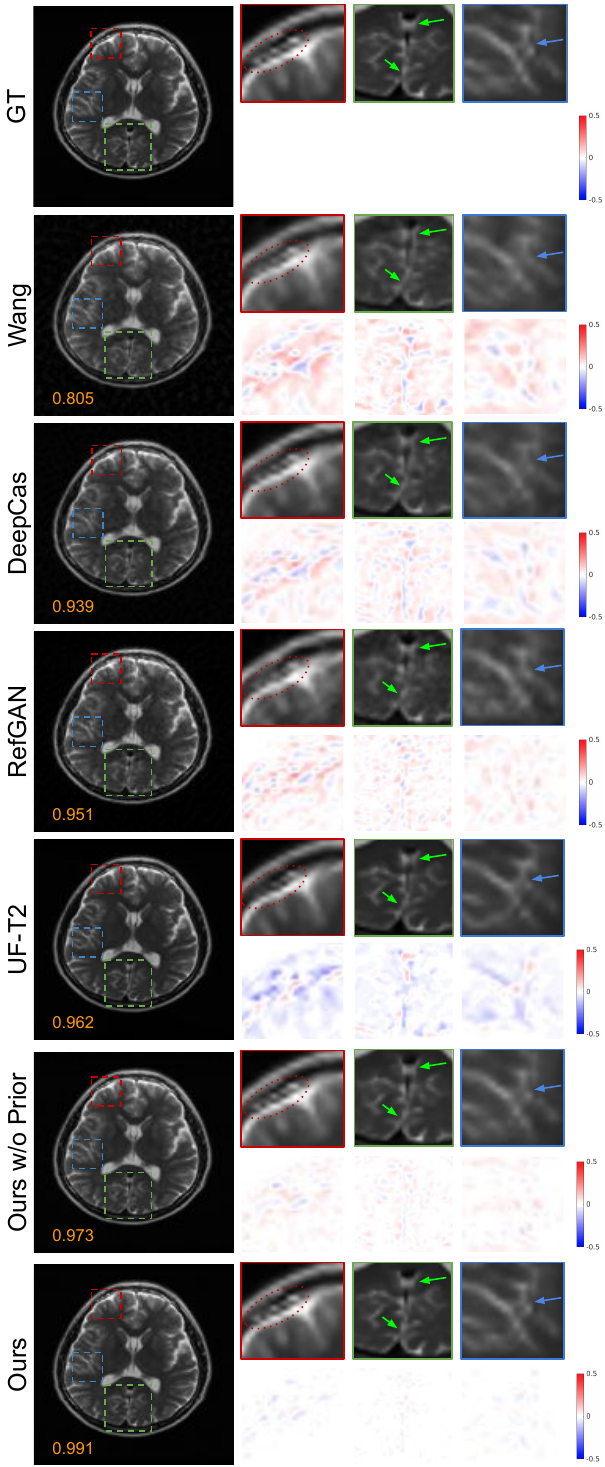}
\caption{Comparison of reconstructions using radial trajectory at an acceleration rate $R=5$. Three enlarged sub-regions and corresponding difference images are shown on the right. SSIM is indicated on the bottom-left.}
\label{fig:comp_image2}
\end{figure}

\begin{figure*}[htb!]
     \centering
    \begin{subfigure}[t]{0.33\textwidth}
        \raisebox{-\height}{\includegraphics[width=\textwidth]{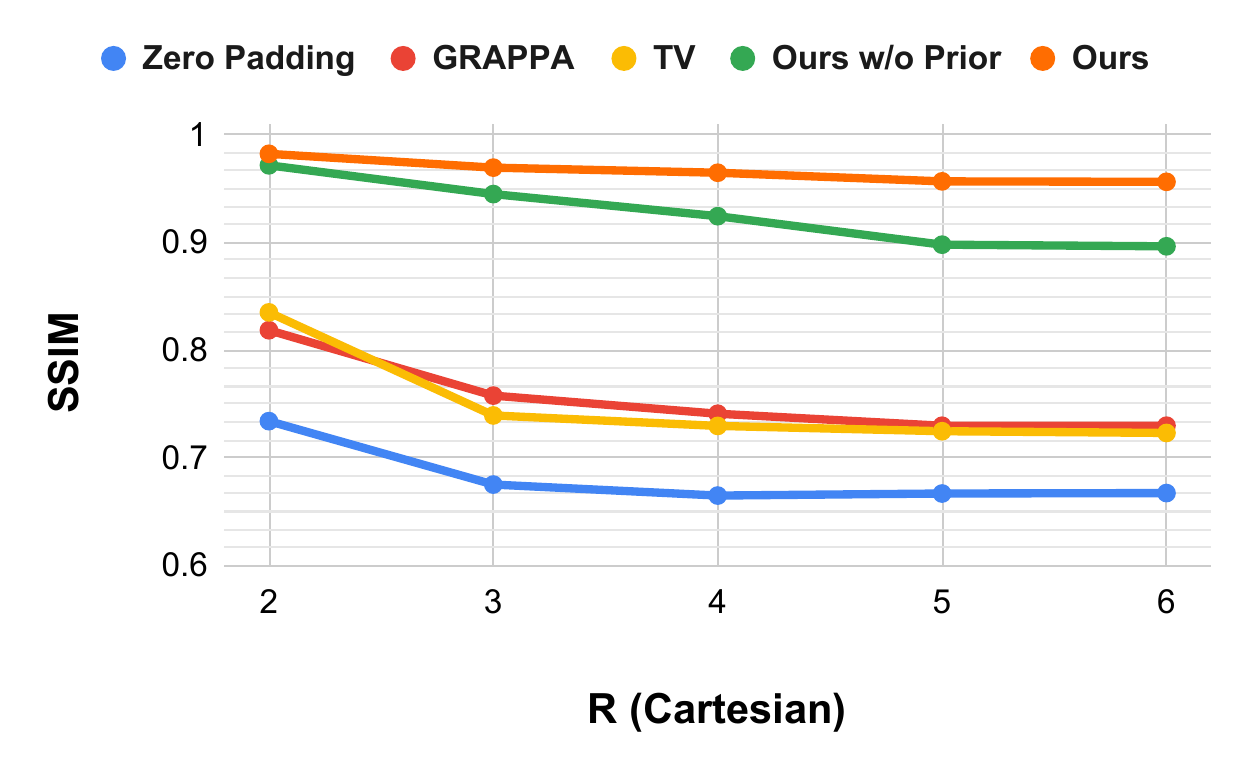}}
    \end{subfigure}
    \hfill
    \begin{subfigure}[t]{0.33\textwidth}
        \raisebox{-\height}{\includegraphics[width=\textwidth]{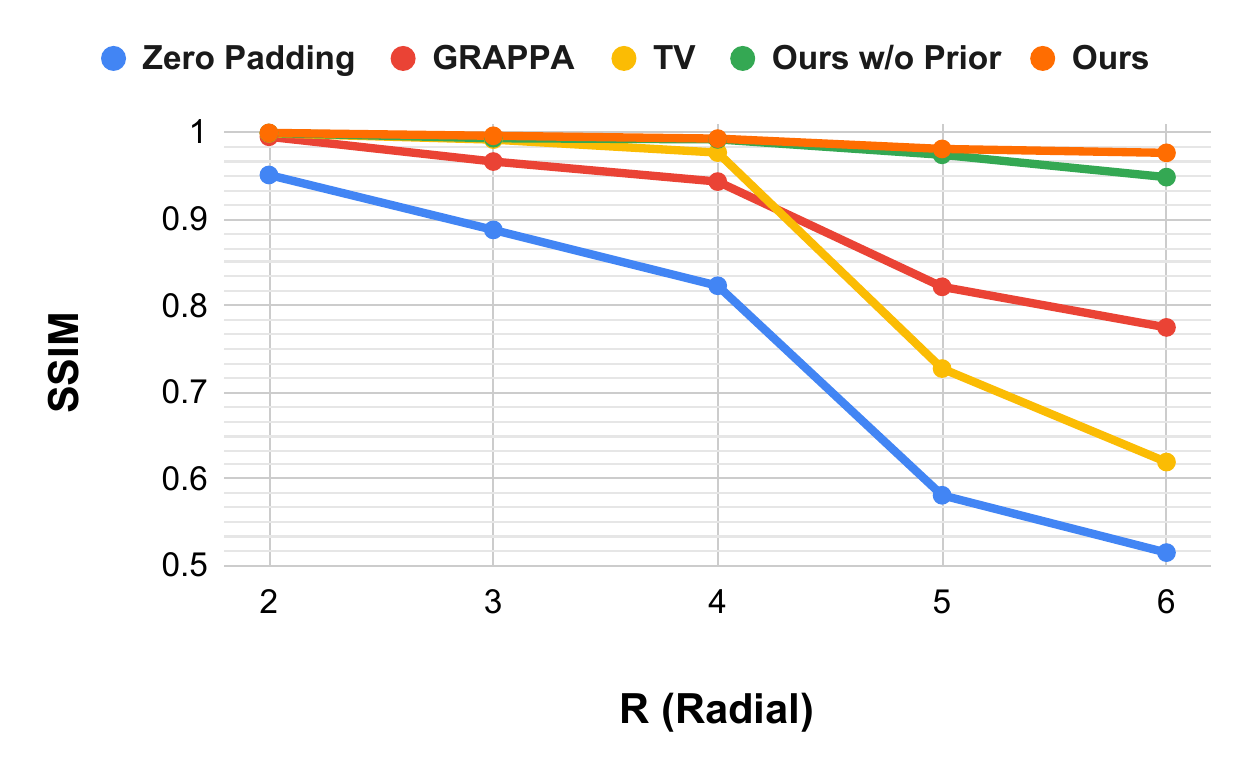}}
    \end{subfigure}
    \begin{subfigure}[t]{0.33\textwidth}
        \raisebox{-\height}{\includegraphics[width=\textwidth]{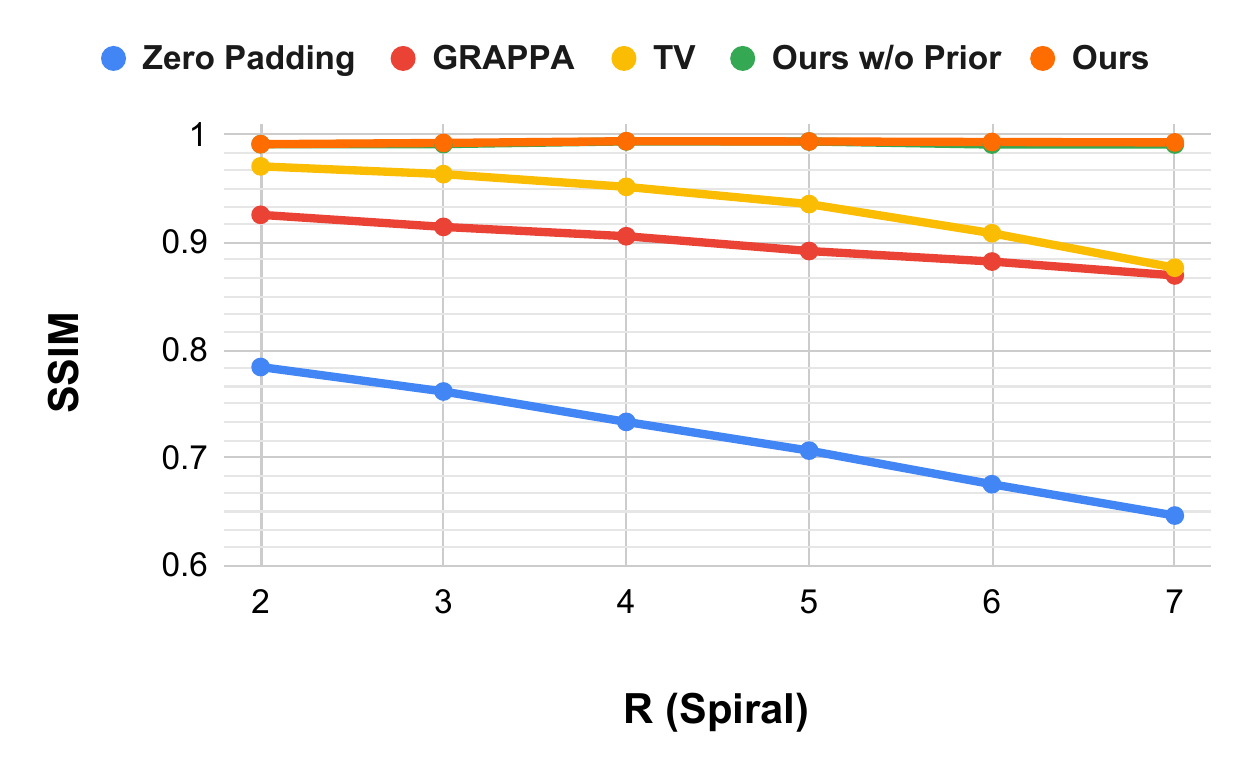}}
    \end{subfigure}
    \caption{Comparison of reconstruction performance at different sampling rate using Cartesian, radial, and spiral patterns.}
    \label{fig:recon_stab}
\end{figure*}

\subsection{Experimental Settings}
\noindent\textbf{Dataset and Training} We acquired an in-house MRI dataset consisting of 20 patients. We scanned each patient using three different protocols (T1, T2, and FLAIR) with full k-space sampled, resulting in three 3D volumes of $320 \times 230 \times 18$ for each patient in both image and k-space domains. 360 2D images are generated for each protocol. We split the dataset patient-wise into training/validation/test sets with a ratio of $7:1:2$. As a result, our dataset consists of 252 training images, 36 validation images, and 72 test images for each protocol. Three different k-space undersampling patterns are examined in our experiments, including Cartesian, radial, and spiral trajectories. Examples of the sampling patterns are illustrated in Figure \ref{fig:comp_image1} (green box). The acceleration factor (R) is set to a value between 1 and 6 for all three patterns, corresponding to acceleration in acquisition time. Using these patterns, randomly undersampled T2/FLAIR and fully-sampled T1 are used as input to our model for training. During training, we also randomly augment the image data by flipping horizontally or vertically, and by rotating at different angles.

\noindent\textbf{Performance Evaluation} 
The final evaluation is performed on our 72 test images. For quantitative evaluation, we evaluate our image reconstruction results using Peak Signal-to-Noise Ratio (PSNR), Structural Similarity Index (SSIM), and Mean Square Error (MSE). For comparative study, we compared our results with the following algorithms in previous works: total variation with penalized compress-sensing (TV-CS) \cite{lustig2007sparse}, GRAPPA \cite{griswold2002generalized}, and four CNN-based algorithms including a sequential convolutional model \cite{wang2016accelerating}, a cascade image restoration model called DeepCas \cite{schlemper2017deep}, a structural refinement GAN model called RefGAN \cite{quan2018compressed}, and a DenseUNet model with complementary T1 image called UF-T2 \cite{xiang2018ultra}. All CNN-based methods are trained using the same settings as to our method. 

\subsection{Results}
\noindent\textbf{Image Quality Evaluation and Comparison}
We evaluated our reconstruction results by computing their SSIM, PSNR, and MSE, where the fully sampled images are used as ground-truth for this calculations. In Table \ref{tab:scan_metrics}, we demonstrated our T2 reconstruction evaluations on three sampling patterns with a high acceleration rate of $R=5$. The first sub-table summarized the image quality when Cartesian-based acceleration is applied. Our DuDoRNet without T1 prior achieved PSNR up to $27.834$ dB, and boosted to $32.511$ dB when T1 prior is given. As compared to the previous state-of-the-art method with T1 prior called UF-T2 \cite{xiang2018ultra}, we improved the reconstruction from $30.594$ dB to $32.511$ dB. Similarly for radial-based acceleration, our DuDoRNet without T1 prior achieved PSNR up to $37.27$ dB, and further boosted to $40.815$ dB when T1 prior is provided. As compared to the best results with PSNR $=35.11$ from RefGAN \cite{quan2018compressed} without T1 prior, our DuDoRNet achieves significantly better results. In the last sub-table, we found our DuDoRNet with spiral pattern yields the best image quality among all sampling patterns and reconstruction methods. Under the spiral pattern, our DuDoRNet without T1 prior achieved PSNR $=48.418$ dB and reinforced to $49.186$ dB when T1 prior is present. The qualitative comparison results with non-CNN based methods are shown in Figure \ref{fig:comp_image1}. As we can see, the reconstructions with zero padding (ZP) at a high acceleration rate create significant aliasing artifacts and lose anatomical details. Non-CNN based methods can improve the reconstruction as compared to ZP, but it is hard to see a significant improvement when a significant level of aliasing artifact is presented. In comparison, our reconstructions are robust to aliasing artifacts and structural loss in the input. The qualitative comparison results with CNN based methods are shown in Figure \ref{fig:comp_image2}. At a high acceleration rate, the CNN based methods achieve better results than non-CNN based methods. Among them, our method restores information in both image and k-space better preserved the important anatomical details, as demonstrated in Figure \ref{fig:comp_image2} with arrows and ellipses. More results on FLAIR reconstruction are summarized in supplemental materials with similar performances.

\noindent\textbf{Reconstruction Stability Evaluation} To evaluate the performance stability at different acceleration rates, we recorded the reconstruction performance by varying $R$ from 2 to 6 on all three patterns. The evaluation results are summarized in Figure \ref{fig:recon_stab}. For the Cartesian pattern, our method can consistently maintain the SSIM to above 0.95 even when an aggressive acceleration rate is used, \ie $R=6$. Due to the large aliasing artifact created from the Cartesian sampling pattern as the acceleration rate increases, the reconstruction performance is more challenging to remain stable without T1 prior, but we were still able to maintain SSIM to above 0.89 and consistently outperforms the other methods for all undersampling rates. For radial pattern, all methods performed approximately the same at low acceleration rates $R<4$. However, for a more aggressive undersampling rate, \ie $R>4$, our DuDoRNet is able to reduce the structural loss by a considerable margin. At $R=6$, Our DuDoRNet with and without T1 prior can consistently maintain the SSIM to above 0.98 and 0.97, respectively. Lastly, we found the best performance stability when the spiral pattern is applied. Our DuDoRNet keeps the SSIM to above 0.99 over the whole undersampling range regardless of T1 prior. Under the same acceleration rate, radial and spiral patterns sample the k-space more uniformly than random Cartesian, thus leading to less aliasing artifact in the initial reconstruction input for the models, as demonstrated in Figure \ref{fig:comp_image1}. As a result, radial and spiral patterns generate less aliasing artifacted input and are able to output more stable reconstruction in our experiments. 

\subsection{Ablation Studies}
\begin{table}[htb!]
\footnotesize
\centering
\caption{Quantitative evaluations for Rec, DD, and DIL components in DuDoRNet. For the average, we use $\dagger$ to indicate if difference were statistically significant ($p < 0.05$) compared to (A).}
\label{tab:component_analysis}
    \begin{tabular}{l||c|c|c||c}
        \hline
        \textbf{SSIM}       & Cartesian  & Radial   & Spiral   & Average  \\
        \hline
        (A) Net-baseline    & $0.839$      & $0.902$    & $0.951$    & $0.897$    \\ [0.025cm]
        \hline
        (B) Net-Rec         & $0.860$      & $0.959$    & $0.973$    & $0.931^\dagger$    \\ [0.025cm]
        \hline
        (C) Net-DD          & $0.851$      & $0.929$    & $0.962$    & $0.914^\dagger$    \\ [0.025cm]
        \hline
        (D) Net-DIL         & $0.844$      & $0.909$    & $0.956$    & $0.903$            \\ [0.025cm]
        \hline
        (E) Net-Rec-DD      & $0.891$      & $0.968$    & $0.989$    & $0.949^\dagger$    \\ [0.025cm]
        \hline
        (F) Net-Rec-DIL     & $0.869$      & $0.962$    & $0.978$    & $0.936^\dagger$    \\ [0.025cm]
        \hline
        (G) Net-DD-DIL      & $0.859$      & $0.935$    & $0.969$    & $0.921^\dagger$    \\ [0.025cm]
        \hline
        (H) Net-Rec-DD-DIL  & $0.898$      & $0.974$    & $0.991$    & $0.954^\dagger$    \\ [0.025cm]
        \hline
    \end{tabular}
\end{table}

Firstly, we evaluated the effectiveness of different components in our DuDoRNet. Without loss of generality, the reconstruction performance is evaluated on $R=5$ for all three patterns. We evaluate three key components, including: dual domain learning (DD), dilated residual dense learning (DIL), and recurrent learning (Rec) without T1 prior. $N_{rec}$ in Rec is set to 5. The component analysis is summarized in Table \ref{tab:component_analysis}. As we can observe, recurrent learning (B) and dual domain learning (C) each improve the performance over the baseline (A) by 0.015, which are more significant than dilated residual dense learning (D). Combining recurrent learning and dual domain learning (E), the performance achieves the largest boost as compared to the other two component combinations (F and G). Our DuDoRNet, equipping all components (H), produces the best reconstruction results. Overall, all three components help DuDoRNet to enhance the performance.

\begin{figure}[htb!]
\centering
\includegraphics[width=0.42\textwidth]{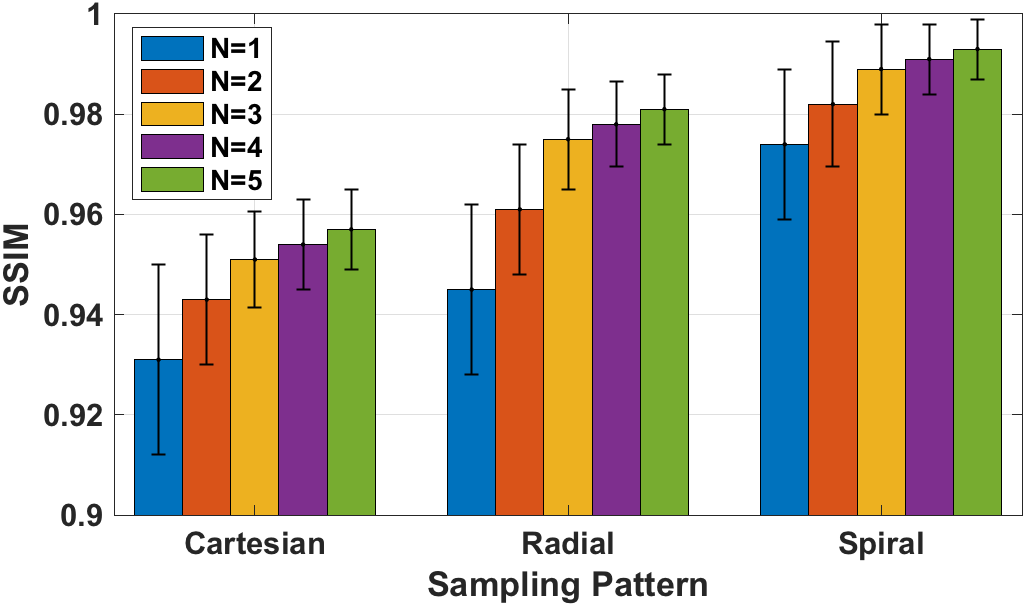}
\caption{The effect of increasing the number of recurrent blocks ($n$) in our DuDoRNet under three sampling patterns.}
\label{fig:as_nrecurrent}
\end{figure}

Secondly, we evaluated the effect of $N_{rec}$ in our DuDoRNet. As shown in Figure \ref{fig:as_nrecurrent}, the reconstruction performance, measured by SSIM, increases monotonically when $N_{rec}$ increases, while the rate of improvement starts to converge after $N_{rec}=3$ for all three patterns. We also found that our DuDoRNet achieves the best performance when the spiral pattern is used even when different $N_{rec}$ is implemented.

\section{Conclusion}
We present a dual domain recurrent network for fast MRI reconstruction with T1 prior embedded. Specifically, we propose to restore both image and k-space domains recurrently through DRD-Nets with large receptive fields. The T1 prior is embedded at each recurrent block to deeply guide restorations for both domains. Extensive experimental results demonstrate that while previous fast MRI methods on single domain for individual protocol have limited capability of directly reducing aliasing artifacts in the image domain, our DuDoRNet can efficiently restore the reconstruction, and the T1 prior can further significantly improve the structural recovery. Future work includes exploring DuDoRNet on pathological MR data and the application to other signal recovery tasks, such as noise reduction and super resolution.

\clearpage
{\small
\bibliographystyle{ieee_fullname}
\bibliography{egbib}
}

\end{document}